\documentclass[aps,epsfig,preprint]{revtex4-1}
\usepackage{soul}
\usepackage[section]{placeins}        %Keep Figure in Section
\usepackage{amsmath}
\usepackage{amssymb}
\usepackage{graphics}
\usepackage{empheq}
\usepackage{caption}
\usepackage{wrapfig}
\usepackage{sidecap}
\usepackage{epsfig}
\usepackage{pdfpages}
\usepackage{float}
\usepackage{subcaption}

\newcommand{\be}{\begin{equation}}
\newcommand{\ee}{\end{equation}}

\newcommand{\bea}{\begin{eqnarray}}
\newcommand{\eea}{\end{eqnarray}}

\begin{document}
\title{ Sub and super-luminar propagation of structures satisfying Poynting like theorem for 
incompressible GHD fluid model depicting strongly coupled dusty plasma 
medium }
\author{Vikram Dharodi}
\author{Amita Das}
\email{amita@ipr.res.in}
\author{Bhavesh Patel }
\author{Predhiman Kaw}
\affiliation{Institute for Plasma Research, Bhat, Gandhinagar - 382428, India }
\date{\today}
\begin{abstract}
The strongly coupled dusty plasma has often been modelled by the Generalized
Hydrodynamic (GHD) model used for representing visco-elastic fluid systems. The
incompressible limit of the model which supports transverse shear wave mode is
studied in detail. In particular dipole structures are observed to emit
transverse shear waves in both the limits of  sub and super luminar propagation,
where the structures move slower and faster than the phase velocity of the shear
waves,  respectively. In the sub - luminar limit the dipole gets engulfed within
the shear waves emitted by itself, which then  backreacts on it and ultimately 
the identity of the structure is lost. However, in the super - luminar limit the
emission appears like a wake from the tail region of the dipole. The dipole,
however,  keeps propagating forward with little damping but minimal distortion
in its form. A Poynting like conservation law with radiative, convective and
dissipative terms being responsible for the evolution of $W$, which is similar
to  `enstrophy' like quantity in normal hydrodynamic fluid systems, has also
been constructed for the incompressible GHD equations. The conservation law is
shown to be satisfied in all the cases of evolution and collision amidst the
nonlinear structures to a great accuracy. It is shown that  monopole structures
which do not move at all but merely radiate shear waves, the radiative term and
dissipative losses solely contribute to the evolution of $W$. The, dipolar
structures, on the other hand, propagate in the medium and hence convection also
plays an important role in the evolution of $W$. 
\end{abstract}
% for revtex4 here maketitle should be written
\pacs{}
\maketitle
%\begin{multicols}{2}
\section{Introduction}
In many physical situations like  earth's lower magnetosphere
\cite{Goertz_1989,Northrop_1992}, planetary atmospheres, cometary tails and
comae, planetary and solar nebulae, asteroids, volcanoes, lightning discharge,
interstellar clouds etc. the usual electron ion plasma is interspersed with some
 heavier mass species (typical mass of $10^{-15}-10^{-10}$ Kg)
\cite{Mendis_2002}. In the plasma environment these heavier dust particles
acquire charges as the electrons stick to its surface. This makes it a third
heavier charged species in the electron - ion plasma system, thereby enabling it
to collectively participate in the dynamics. Such plasma are termed as   ``dusty
plasmas" \cite{shukla2010introduction}. This medium has immense applications 
 ranging from industrial study \cite{laifa_Industrial_2002,fridman2004plasma} to
space physics \cite{Merlino2004, Thomas1996_nature_Crystallization}.
Furthermore, as the dusty plasma can be easily prepared to be in a strong
coupling regime at normal densities and room temperature \cite{chu_lin_1994,
Thomas1994Crystallization} it provides a unique opportunity of understanding a
strongly coupled state of matter (an area of prime importance)  in a simplified
set up \cite{fortov_book,bannasch_13,horn_91}. 
%~~~~~~~~~~~~~~~~~~~~~~~~~~~~~~~~~~~~~~~ 
\paragraph*{}
%~~~~~~~~~~~~~~~~~~~~~~~~~~~~~~~~~~~~~~~
The strongly coupled state of the dusty plasma medium has been studied often by
a simplified visco - elastic fluid description using the Generalized
Hydrodynamic fluid model
\cite{Kaw_Sen_1998,Kaw_2001,banerjee2010viscosity,Sanat_kh_sronge_2012,
Sanat_jpp_2014,tiwari2015turbulence,Ghosh2011_nonlinearwave,Janaki2010vortex,
shukla2001dust}.
The strong coupling nature of the system is mimicked here by a relaxation time
parameter. The system is assumed to retain memory and hence act like an elastic
medium for time scales shorter than the relaxation time $\tau_m$
\cite{frenkel_kinetic,Kaw_Sen_1998,Kaw_2001, berkovsky_1992, Ichimaru_1987}. At
longer time scales the usual hydrodynamic nature is supposed to take over.  The
presence of elasticity enables the medium to support transverse shear modes
which are essentially damped in the context of hydrodynamic fluids. This model
has been successful in predicting the dispersion characteristics of the
transverse shear wave (TSW) in the medium \cite{Kaw_Sen_1998,Kaw_2001}, which
have been experimentally demonstrated \cite{Pramanik_2002}. The mode dispersion
has also been obtained within the Molecular Dynamics (MD) simulations which
treat the dust particles (screened by the background electrons and ions)  
as interacting through Yukawa potential \cite{Ohta_Hamaguchi_2000}.
%~~~~~~~~~~~~~~~~~~~~~~~~~~~~~~~~~~~~~~~ 
\paragraph*{}
%~~~~~~~~~~~~~~~~~~~~~~~~~~~~~~~~~~~~~~~
To isolate the features associated with the transverse shear modes from the
compressible acoustic waves which is also supported by this medium, we consider
here an incompressible limit of the GHD model.  The  details of the  governing
equations of the incompressible GHD (i-GHD) equations have been presented in
section II. 
%~~~~~~~~~~~~~~~~~~~~~~~~~~~~~~~~~~~~~~~ 
\paragraph*{}
%~~~~~~~~~~~~~~~~~~~~~~~~~~~~~~~~~~~~~~~
We then focus on the study of two aspects associated with this set of governing
equations. The conservation laws satisfied by any evolution equation helps
provide important insights on the evolutionary behavior of any system. Keeping
this in view, the i-GHD set of equations were analyzed for a possible
construction of such laws. We obtain a kind of Poynting theorem for an
``enstrophy" like integral  associated with  the  i-GHD system. The mean
square integral quantity is shown to   decay due to  dissipation and  through
convection and emission of waves. The validity of this theorem is then 
numerically verified in several contexts in subsequent sections.
%~~~~~~~~~~~~~~~~~~~~~~~~~~~~~~~~~~~~~~~ 
\paragraph*{}
%~~~~~~~~~~~~~~~~~~~~~~~~~~~~~~~~~~~~~~~
The second aspect we focus on is related to the evolution and interaction of
coherent vortex structures. In our previous work \cite{Dharodi_2014}, it was
shown  that in contrast to Newtonian fluids, visco-elastic fluid (described by 
i-GHD model) supports the emission of transverse shear waves from the rotating
vorticity patches. The phase propagation speed was observed to match the
theoretical prediction of $\sqrt{\eta/\tau_m}$  as predicted by Kaw {\it et al.}
\cite{Kaw_Sen_1998}. The other important structure which has been studied
extensively in the context of Newtonian fluids has the dipolar symmetry and is
essentially a structure which forms when two unlike signed monopoles are brought
together, they form a dipolar structures which propagates in a stable fashion
along its axis in hydrodynamic fluids. The evolution of dipolar structures are
studied in detail here in the context of i-GHD model and has been presented in
section IV. It is shown that the dipoles also do emit transverse shear waves as
expected. However, there are two interesting cases in the simulation. When the
dipole moves slower than the phase velocity of the emitted waves (sub - luminar)
 it gets totally engulfed within the propagating waves which react and distort
the original dipole structure pretty soon. In the other limit when the dipoles
move faster than phase velocity of the transverse shear  waves in the
medium, the TSW are emitted from the tail of the structure in the form of a
wake. The dipole, however, continues to move as a stable entity with a conical
wake of waves trailing behind it. We have also considered the collisional
interaction amidst two oppositely moving dipoles. Here too they behave like
hydrodynamic fluid, they exchange partners and move in the orthogonal direction
in the super-luminar cases. We have then numerically verified the validity of
the Poynting like theorem for the i-GHD set of equations for all these cases of
propagation.
%~~~~~~~~~~~~~~~~~~~~~~~~~~~~~~~~~~~~~~~ 
\paragraph*{}
%~~~~~~~~~~~~~~~~~~~~~~~~~~~~~~~~~~~~~~~ 
The paper has been organized as follows. Section~\ref{Governing_Equations}
contains the details of the governing equations. In section~\ref{energych2} we
derive analytically a Poynting like conservation equation for our i-GHD system. 
In section~\ref{dipole_evo_ch2}, various cases of dipole evolution and
interaction have been presented  showing the influence of the emitted 
transverse shear waves on the sanctity of these structures. In
section~\ref{num_veri_poy_ch2} we present the simulation studies which confirm
the validity of the Poynting like theorem and help identify the dominant
mechanism of the decay for  the `enstrophy' like integral of the system. 
Finally, section ~\ref{discussionch2} contains the discussion and conclusion.
%~~~~~~~~~~~~~~~~~~~~~~~~~~~~~~~~~~~~~~~
%
\section{Governing Equations}
\label{Governing_Equations}
\paragraph*{}
The strongly coupled dusty plasma system has been analysed with the help of 
coupled set of  continuity, GHD momentum and Poisson's  equation, both 
analytically as well as numerically to a great extent in  past studies 
\cite{Kaw_Sen_1998,Kaw_2001,banerjee2010viscosity,Sanat_kh_sronge_2012,
Sanat_jpp_2014,tiwari2015turbulence,Ghosh2011_nonlinearwave,Janaki2010vortex,
shukla2001dust}.
% \cite{Kaw_Sen_1998,
% Kaw_2001,Sanat_aip_2012,Sanat_singular_2012,Sanat_njp_solitons_2012,
%  Sanat_kh_weak_2012,Sanat_kh_sronge_2012,Sanat_jpp_2014,tiwari2015turbulence,
% Janaki2011jeans,banerjee2010viscosity, Ghosh2011_nonlinearwave,
% banerjee2010viscosity, Janaki2010vortex,shukla2001dust}.
The set of these equations permit both the existence of incompressible
transverse shear and compressible longitudinal modes. We choose in this paper to
concentrate on the  incompressible features of this system by separating out the
compressibility effects altogether. For this purpose, the incompressible limit
of the GHD (i-GHD) set of equations have been obtained. In the incompressible
limit the Poisson equation is replaced by the quasi neutrality condition and
charge density fluctuations are ignored. The derivation of this reduced equation
is discussed in detail in our earlier paper \cite{Dharodi_2014} along with the
procedure of its  numerical implementation and validation.
%~~~~~~~~~~~~~~~~~~~~~~~~~~~~~~~~~~~~~~~
\paragraph*{}
%~~~~~~~~~~~~~~~~~~~~~~~~~~~~~~~~~~~~~~~ 
The momentum equation for GHD  of strongly coupled homogeneous dusty plasma
{~\cite{Dharodi_2014}} in the incompressible limit can be written as: 
{
\begin{equation}\label{eq:in_momentum}
 \left[1 + \tau_m \left(\frac{\partial}{\partial t}+\vec{v}_d  \cdot \nabla
\right)\right]  \left[{\frac{\partial \vec{v}_d } {\partial t}+\vec{v}_d  \cdot
\nabla\vec{v}_d } + \frac{\nabla P}{n_d} - \nabla \phi \right]  = \eta \nabla^2
\vec{v}_d  
  \end{equation}
  }
 $\eta $ is now termed as the kinematic viscosity. The time is normalised by
inverse of dust plasma frequency {$\omega_{pd}^{-1}$} and the length is
normalised by plasma Debye length {$\lambda_d$}. The Standard Navier Stokes
equation can be achieved from Eq.~(\ref{eq:in_momentum}) by taking $\tau_m =0$.
For numerical convenience we can  split the Eq.~(\ref{eq:in_momentum}) in terms
of following two coupled equations. 
 \begin{eqnarray}\label{eq:vorticity4}
{{\frac{\partial \vec{v}_d } {\partial t}+\vec{v}_d  
\cdot \nabla\vec{v}_d }+ \frac{\nabla P}{n_d} -\nabla \phi }={\vec \psi} 
\end{eqnarray}
\begin{eqnarray}\label{eq:psi4}
\frac{\partial {\vec \psi}} {\partial t}+\vec{v}_d \cdot \nabla{\vec \psi}=
{\frac{\eta}{\tau_m}}{\nabla^2}{\vec{v}_d }-{\frac{\vec \psi}{\tau_m}}
\end{eqnarray}
%~~~~~~~~~~~~~~~~~~~~~~~~~~~~~~~~~~~~~~~
Thus the Eq.~(\ref{eq:in_momentum}) has  now been  expressed as a set of two
coupled convective equations. The gradient terms are eliminated by taking the
curl of Eq.~(\ref{eq:vorticity4}) which yields an equation for the evolution of
the vorticity field. So the  coupled set of
Eqs.(\ref{eq:vorticity4})-(\ref{eq:psi4}) have been recast in the following
form. 
 \begin{eqnarray}\label{eq:vort_num}
\frac{\partial{\vec \xi}} {\partial t}+\vec{v}_d \cdot \nabla{{\vec \xi}}={\vec
\nabla}{\times}{\vec \psi}
\end{eqnarray}
\begin{eqnarray}\label{eq:psi_num}
\frac{\partial {\vec \psi}} {\partial t}+\vec{v}_d \cdot \nabla{\vec \psi}=
{\frac{\eta}{\tau_m}}{\nabla^2}{\vec{v}_d }-{\frac{\vec \psi}{\tau_m}}
\end{eqnarray}
\paragraph*{}
 Equations (\ref{eq:vort_num}) and (\ref{eq:psi_num}) are a coupled set of
closed equations for a visco-elastic fluid in the incompressible limit. These
equations would be referred as i-GHD model equations henceforth in the manuscript. 
 Here, $ {\vec \xi}={\vec\nabla}{\times}{\vec{v}_d} $ (here $ {\vec \xi}$
normalised with $\omega_{pd}$) is the vorticity. It should be noted that in
this particular limit there is nothing specific which is suggestive of the fact
that the system corresponds to a strongly coupled dusty plasma medium. Thus, the
results presented in this paper (based on GHD model) would in general be
applicable to any visco-elastic medium and need not  be restricted to the
strongly coupled dusty plasma medium. The reduced set of equation, not only
caters to the strongly coupled incompressible dusty plasma medium but is
relevant for any other incompressible visco -elastic system. 
%%However, the dusty plasma medium being one of the most 
%%easily amenable strongly coupled system experimentally, it is likely that some 
%%of the predicted features may find experimental support in near future. 
%%%%%%%%%%%%%%%%%%%%%%%%%%%%%%%%%%%%%
\section{A Poynting like Theorem for the coupled set of i-GHD}
\label{energych2}
\paragraph*{}
An interesting Poynting like theorem can be obtained for the i-GHD model. Such
conservation equations are in general a powerful tool  for any system. They
provide interesting physical insights for the system and can also be employed
for validating as well as discerning the accuracy of any numerical program. 

\paragraph*{}
Taking the dot products with respect to $\vec{\xi}$ and $\vec{\psi}$ for
equations~(\ref{eq:vort_num}) and (\ref{eq:psi_num}) respectively we obtain: 
\begin{eqnarray}\label{eq:vort_poy}
{\frac{1}{2}}\frac{\partial{\xi_z^2}} {\partial t}
+{\vec \xi}{\cdot}{\left(\vec{v} \cdot \vec \nabla\right){{\vec \xi}}}
={\vec \xi}{\cdot}{{\vec \nabla}{\times}{\vec \psi}}   
\end{eqnarray}
\begin{eqnarray}\label{eq:psi_poy}
{\frac{1}{2}}\frac{\partial {{\psi}^2}} {\partial t}
+{{\vec \psi}}{ \cdot }\left(\vec{v} \cdot \vec \nabla\right){\vec \psi}
={{\vec\psi}\cdot}{\frac{\eta}{\tau_m}}{\nabla^2}{\vec{v}}-{\frac{{{\psi^2}}}{
\tau_m}}        
\end{eqnarray}
It should be noted that the vorticity vector $\vec \xi$ in the 2-D geometry has
only $\hat{z}$ component. We have the following vector relations. 
\begin{equation}
{{\vec \xi}}{\cdot}\left(\vec{v} \cdot \vec \nabla\right){\vec \xi}
={\nabla}{\cdot}({\vec{v}}{\frac{\xi_z^2}{2}})    \nonumber       
\end{equation}
 \begin{equation}
{{\vec \psi}}{\cdot}\left(\vec{v} \cdot \vec \nabla\right){\vec \psi}
={\nabla}{\cdot}({\vec{v}}{\frac{\psi^2}{2}}) \nonumber 
\end{equation}
\begin{equation}
{{\vec \psi}}{\cdot}{\nabla^2}{\vec v}=-{\xi_{z}}{\cdot}({{\vec
\nabla}{\times}{\vec
\psi}})- {\vec \nabla}{\cdot}({\xi_{z}}{{\times}{\vec \psi}})   \nonumber  
\end{equation}
Using the first vector relation and multiplying Eq.~(\ref{eq:vort_poy}) by
${\eta}/{\tau_m}$ we have 
\begin{eqnarray}\label{eq:vorticity2r}
{\frac{1}{2}}{\frac{\eta}{\tau_m}}\frac{\partial{\xi^2_{z}}} {\partial t}
+{\frac{\eta}{\tau_m}}{\nabla}{\cdot}({\vec {v}}{\frac{\xi_z^2}{2}})
={\frac{\eta}{\tau_m}}{\xi}_{z}{\cdot}{{\vec \nabla}{\times}{\vec \psi}}   
\end{eqnarray}
The other two vector relations are used in Eq.~(\ref{eq:psi_poy}) to obtain : 
\begin{eqnarray}\label{eq:psi2}
{\frac{1}{2}}\frac{\partial {{\psi}^2}} {\partial t}
+{\nabla}{\cdot}({\vec v}{\frac{\psi^2}{2}}) =
-{\frac{\eta}{\tau_m}}{\xi_{z}}{\cdot}({{\vec \nabla}{\times}{\vec \psi}})-
 {\frac{\eta}{\tau_m}}{\vec \nabla}{\cdot}({\xi_{z}}{{\times}{\vec \psi}})
-{\frac{{{\psi^2}}}{\tau_m}}       
\end{eqnarray}
Now summing  equations ~(\ref{eq:vorticity2r}) and ~(\ref{eq:psi2}), we get
\begin{empheq}[]{align}\label{eq:energy_sum1}
{\frac{\partial}{\partial
t}}{\left(\frac{\psi^2}{2}+{\frac{\eta}{\tau_m}}\frac{\xi_z^2}{2}
\right)}
+{\vec \nabla}{\cdot}{\frac{\eta}{\tau_m}}({\omega_{z}}{{\times}{\vec \psi}})
+{\nabla}{\cdot}{\vec v}\left({{\frac{\psi^2}{2}}}
+{\frac{\eta}{\tau_m}}{\frac{\omega_z^2}{2}}\right)
=-{\frac{\psi^2}{\tau_m}}
\end{empheq}
Clearly, the form of Eq.~(\ref{eq:energy_sum1}) is that of the Poynting like
equation
\begin{empheq}[box=\fbox]{align}\label{eq:denergy}
{\frac{\partial W}{\partial t}}+{\nabla}{\cdot}{\vec
S}+{\nabla}{\cdot}{(T\vec v)}+P_d =0
\end{empheq}
with following identifications. 
$W\equiv\left(\frac{\psi^2}{2}+{\frac{\eta}{\tau_m}}\frac{\xi_z^2}{2}
\right)$, ${\vec
S}\equiv{\frac{\eta}{\tau_m}}({\xi_{z}}{{\times}{\vec \psi}})$, 
$P_d\equiv{\frac{\psi^2}{\tau_m}}$ 
 and $T\equiv\left(\frac{\psi^2}{2}+{\frac{\eta}{\tau_m}}\frac{\xi_z^2}{2}
\right)$. 
This shows that the rate of change of $W$ depends on dissipation through $P_d$
in the medium, a convective and radiative flux of $T \vec{v}$ and $\vec{S}$
respectively. The radiative Poynting flux,  as we would see later is associated
with the emission of transverse shear waves in the medium.
Equation~(\ref{eq:energy_sum1}) can also be recast in the following integral
form: 
\begin{eqnarray}
{{{\frac{\partial}{\partial t}}}{\int_V}Wdv}+{\oint_{S}}{\vec S{\cdot}d{\vec
a}}+{\oint_{S}}{T~\vec v{\cdot}d{\vec a}}=-{\int_V}{P_d}{dv}
\end{eqnarray}
which corresponds to 
 \begin{equation}
%   \label{eq:integral_equ} 
{\frac{\partial}{\partial t}}
{\int_{V}}{\left(\frac{\psi^2}{2}+{\frac{\eta}{\tau_m}}\frac{\xi_z^2}{2}
\right)}dv+{\frac{\eta}{\tau_m}}{\oint_{S}}({\xi_{z}}{{\times}{\vec
\psi}}){\cdot}d{\bf{a}}+{\oint_{S}}\left(\frac{\psi^2}{2}+{\frac{\eta}{\tau_m}}
\frac{\xi_z^2}{2}\right){\vec v}{\cdot}d{\bf{a}} 
=-{\int_{V}}{\frac{\psi^2}{\tau_m }}dv
 \end{equation}
or
\begin{multline}\label{eq:integral_equ} 
\underbrace{{\frac{\partial}{\partial t}}
{\int_{V}}{\left(\frac{\psi^2}{2}+{\frac{\eta}{\tau_m}}\frac{\xi_z^2}{2}
\right)}dv}_{\text{\bf dWdt}}= \\
-\underbrace{{\frac{\eta}{\tau_m}}{\oint_{S}}({\xi_{z}}{{\times}{ \vec
\psi}}){\cdot}d{\bf{a}}}_{\text{
\bf S}}-\underbrace{{\oint_{S}}\left(\frac{\psi^2 }{2 }+{\frac{
\eta}{\tau_m}}\frac{\xi_z^2}{2}\right){\vec v}{\cdot}d{\bf{a}}}_{\text{\bf T}} 
-\underbrace{{\int_{V}}{\frac{\psi^2}{\tau_m }}dv}_{\text{\bf P}}  
 \end{multline}
 
 It is interesting to physically analyse each of the terms. The contributions
to $W$ arises from two mean square integrals. While $\xi_z$ can easily be
identified with the $z$ component of  vorticity which is typically conserved in
2-dimension Newtonian fluids, the quantity $\vec{\psi}$ relates to the strain
created in the elastic medium by the time varying velocity fields. Thus $W$ is
the sum of square integrals of vorticity and the velocity strain.  The radiation
 term $S$ conatins the integral of the cross product of $\xi_z \hat{z}$ and
$\vec{\psi}$. This term is like a Poynting flux for the radiation corresponding
to the transverse shear waves. A comparison with electromagnetic light waves
where $\vec{E} \times \vec{B}$ acts as a radiation flux shows that the
corresponding  two fields here are $\xi_z \hat{z}$ and $\vec{\psi} $. The
equation also clearly shows that the other mechanism causing the change in $W$
is through convection (which would vanish if the velocity normal to  the
boundary region is zero) and the viscous dissipation through $\eta$. 
%~~~~~~~~~~~~~~~~~~~~~~~~~~~~~~~~~~~~~~~
\paragraph*{}
%~~~~~~~~~~~~~~~~~~~~~~~~~~~~~~~~~~~~~~~  
Later, in section \ref{num_veri_poy_ch2} simulation studies have been performed
showing that the   theorem is satisfied in remarkable accuracy for even the most
 complicated  simulation cases that have been considered by us. It also helps
identify prominent mechanism of  decay in $W$ is various scenarios.  

%%%%%%%%%%%%%%%%%%%%%%%%%%%%%%%%%%%%%%%%%%%%%%%%%%%%%%%%%%%%55
\section{ Evolution of localized structures }
\label{dipole_evo_ch2}
\paragraph*{}
It was shown in one of our earlier works \cite{Dharodi_2014} that a monopolar
rotating vortex in the context of GHD emits transverse shear waves. The
amplitude of the vorticity associated with the emission was observed to fall
off radially as $1/\sqrt{r}$. Thus $\bf S$ in Eq.~(\ref{eq:integral_equ}) would
be finite even at infinity, thereby qualifying as a radiative flux term.  
%~~~~~~~~~~~~~~~~~~~~~~~~~~~~~~~~~~~~~~~
\paragraph*{}
%~~~~~~~~~~~~~~~~~~~~~~~~~~~~~~~~~~~~~~~
The radiative emission from monopoles have the same circular symmetry of the
structure. In this regard, therefore, it in interesting to study the emission of
waves from non symmetric structures. The dipoles are ideal as not only they have
broken circular symmetry, but since they propagate axially, cases wherein their
speed exceeds (super - luminar) and/or is slower (sub - luminar) than the phase
velocity of the transverse shear wave can be investigated. We present numerical
evolution studies for these two distinct cases.  We also carry out studies to
understand the collisional interaction of oppositely propagating dipole
structures. 
%%%%%%%%%%%%%%%%%%%%%%%%%%%%%%%%%%%% 
\subsection{Evolution of dipole structures}
\begin{figure}[!h]
        \centering         
\includegraphics[width=\textwidth]{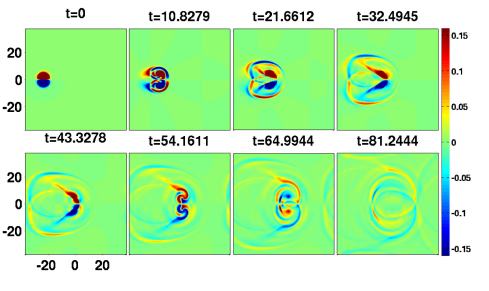}
         \caption{Evolution of sub - luminar dipole in time for visco-elastic
fluid of ${\Omega_0}$=3.5 with the coupling parameters $ {\eta}=5,
{\tau_m}=20$.}
               \label{fig:dipole_sub_520_zita3p5}
       \end{figure}

\begin{figure}[!h]
        \centering         
\includegraphics[width=\textwidth]{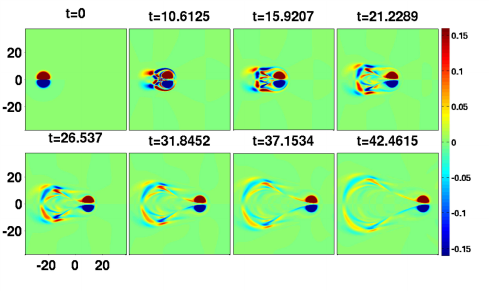}
         \caption{Evolution of super - luminar dipole in time for
visco-elastic fluid of ${\Omega_0}$=5 with the coupling parameters
${\eta}=5, {\tau_m}=20$.}
               \label{fig:dipole_wake_520_zita5}
       \end{figure}

\begin{figure}[!h]
        \centering         
\includegraphics[width=\textwidth]{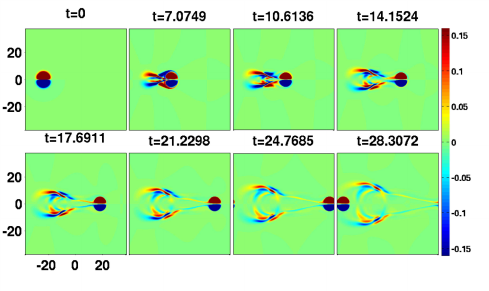}
         \caption{Evolution of super - luminar dipole in time for
visco-elastic fluid of ${\Omega_0}$=7.5 with the coupling parameters $
{\eta}=5, {\tau_m}=20$.}
               \label{fig:dipole_sup_520_zita7p5}
       \end{figure}

\begin{figure}[!h]
        \centering         
\includegraphics[width=\textwidth]{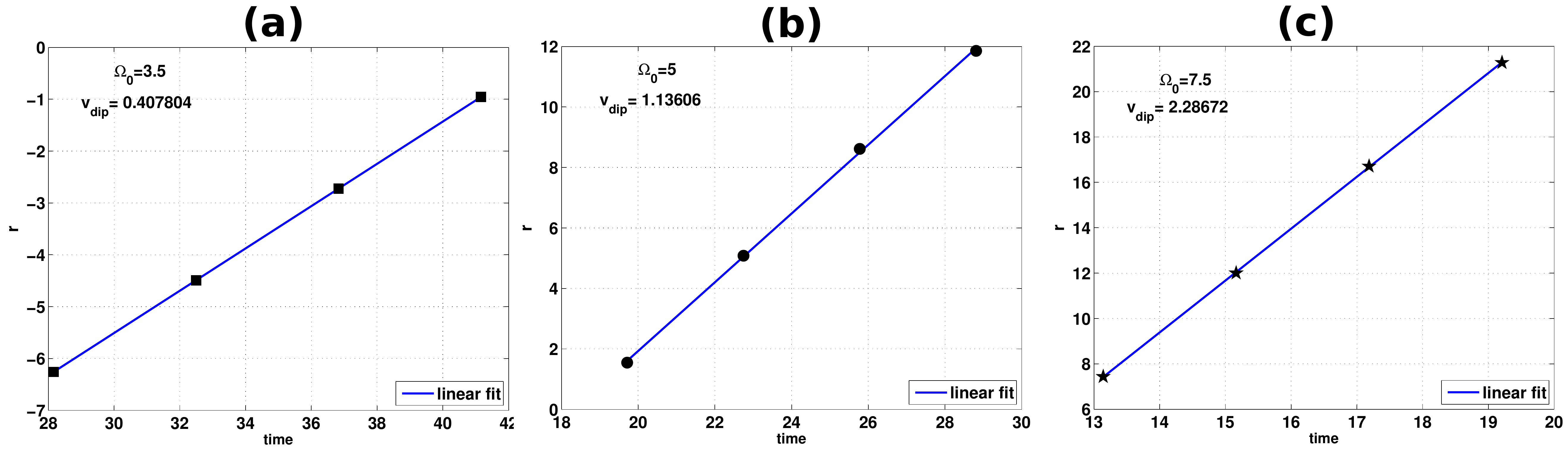}
              \caption{ Position of maximum of dipole amplitude at different
time steps along axial direction with parameter values (a) $\Omega_{0}$=3.5
{$(\blacksquare$)}; $v_{dip}=0.4{<}v_p$ corresponds to sub - luminar dipole, (b)
$\Omega_{0}$=5 ({$\color{black}{ \bullet}$)}; $v_{dip}=1.14 {>}v_p$ corresponds
to super - luminar dipole, and (c) $\Omega_{0}$=7.5  {$ (\color{black}
\bigstar$)}; $v_{dip}=2.29{>}v_p$ is also corresponds to super - luminar dipole,
where $v_{dip}$ is the corresponding axial velocity of dipole related to
$\Omega_{0}$ and blue line is linear fitted curve.}
               \label{fig:time_amp_520_all}
       \end{figure}

\paragraph*{}
When two monopoles rotating in opposite directions (i.e. unlike sign vortices),
are  brought close   they take a shape of a dipole which propagates  along the
direction of its  axis as a single stable entity in the context of Newtonian
fluids. For present case the dipole vorticity profile is given by,
${\xi_{0}(x,y,t=0)}={\Omega_0}(y-y_c)exp\left(-\left({
\left(x-x_c\right)^2+(y-y_c)^2}\right)/{a^2_c}\right)$.
Here $a_c$ is the vortex core radius and numerical simulation has been carried
out for ${a_c}$=2.5, ${x_c}=-24$ and ${y_c}$=0. To satisfy the incompressible
condition, the corresponding velocity is calculated by using the Poisson's
equation $ {\nabla^2}{\vec{v}_0}=-{\vec \nabla}{\times}{\vec \xi}_0$.
 %%%%%%%%%%%%%%%%%%%%%%%%%%%%%%%%%%%%%%%%%%%%%%%%%%%%%
 From Fig.~\ref{fig:dipole_sub_520_zita3p5} to
Fig.~\ref{fig:dipole_sup_520_zita7p5} we show three different cases of
simulations. The simulation region is a square box of length $ 24\pi$ units with
periodic conditions. The value of the  parameters $\eta =5$ and $\tau_m = 20$
has been chosen to be  same for all these three cases. The transverse shear wave
emerge with the phase velocity $v_p =\sqrt{\eta/\tau_m}= 0.5$ for the parameters
chosen for these simulations. The three cases (a, b and c)  have different
amplitude of vorticity ( e.g. $\Omega_{0}$ of $3.5$, $5$ and $7.5$ respectively)
which makes them move with increasing axial speeds. The axial speed of the
dipoles  $v_{dip}$ turns out to be  $0.4{<}v_p$, $1.14{>}v_p$, and $2.29{>}v_p$,
for cases (a), (b) and (c) respectively. This is evidenced from the plot of 
traversed distance vs. time plot for the peak of the structure shown in 
Fig.~\ref{fig:time_amp_520_all} for the three cases.  Clearly, while case (a)
corresponds to a sub - luminar speed of the dipole, (b) and (c) are super -
luminar.
%~~~~~~~~~~~~~~~~~~~~~~~~~~~~~~~~~~~~~~~
\paragraph*{}
%~~~~~~~~~~~~~~~~~~~~~~~~~~~~~~~~~~~~~~~ 
For all these three cases the dipole emits transverse shear wave. However, in case (a)
the dipole gets completely engulfed into the emissions. These emissions then
react on the original structure and the distortions increases with time. It
should also be noted  that the emission from each of the lobes gets
significantly impeded by that of the other as a result of which the emission
profile is no longer symmetrically centered around each of the lobe. The wave
emission from each lobe pushes the other lobe as a result of which the tail end 
of the two lobes can be seen to get pushed away significantly apart. This
increased separation amidst the two lobes as well as the continuous sapping of
the strength of the dipole due to wave emission appears to impact the dipole
propagation speed which can be observed to slow  down as shown in the figure 
Fig.~\ref{fig:dipole_sub_520_zita3p5} and there is also a considerable
distortion in the structure. At later times the lobes have been observed to
rotate and newer structures emerge, resulting in a reformed weak dipole with
reversed polarity  propagating backwards to its original direction. In the
process of such a reformation the merging of like sign vorticity patches and
emission patterns play and important role. Ultimately the identity of the
original dipole structure gets completely lost.
%%%%%%%%%%%%%%%%%%%%%%%%%%%%%%%%%%%%%%%%%%%%%%%%%%%%%%%%%%%%%  
       \FloatBarrier    
%%%%%%%%%%%%%%%%%%%%%%%%%%%%%%%%%%%%%%%%%%%%%%%%%%%%%%%%%%%%%%       
 In the other two cases, however, the dipole structure continues to maintain
its identity. The wave emission in these cases (because of the super-luminar
velocity of dipole) remains confined to a conical spatial regime at the tail.
The radiation merely separates the tail region of the dipole somewhat.
%~~~~~~~~~~~~~~~~~~~~~~~~~~~~~~~~~~~~~~~
\paragraph*{}
In Fig.~\ref{fig:dipole_wake_520_zita5}, we show the dipole evolution with
more strength i.e. larger velocity amplitude. In this case we restrict speed of
dipole not too larger that it leaves behind wake structures. It can be observed
that there is wake type structure formation.
%%%%%%%%%%%%%%%%%%%%%%%%%%%%%%%%%%%%%%%%%%%%%%%%%%%%%%%%%%%%%  
       \FloatBarrier      
%%%%%%%%%%%%%%%%%%%%%%%%%%%%%%%%%%%%%%%%%%%%%%%%%%%%%%%%%%%%%% 
In Fig.~\ref{fig:dipole_sup_520_zita7p5}, we consider the case of dipole moving
with more strength than both former cases. The dipole get out of the cage of
wake structures.
%%%%%%%%%%%%%%%%%%%%%%%%%%%%%%%%%%%%%%%%%%%%%%%%%%%%%%%%%%%%%
       \FloatBarrier       
%%%%%%%%%%%%%%%%%%%%%%%%%%%%%%%%%%%%%%%%%%%%%%%%%%%%%%%%%%%%%%%% 
       \FloatBarrier    
%%%%%%%%%%%%%%%%%%%%%%%%%%%%%%%%%%%%%%%%%%%%%%%%%%%%%%%%%%%%%%  
With increasing speed  of the dipole  the cone  angle between which the TSW
radiation  is confined is observed to reduce. 
%%%%%%%%%%%%%%%%%%%%%%%%%%%%%%%%%%%%%%%%%%%%%%%%%%%%%%%%%%%%%% 
\subsection{Head on collision between dipoles}
\begin{figure}[!h]
        \centering         
\includegraphics[width=\textwidth]
{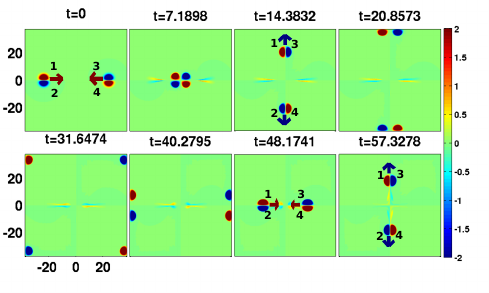}
         \caption{Dipole-dipole head to head collision for hydrodynamic fluid
with ${\Omega_{01}}$=${\Omega_{02}}$=7.5.}
               \label{fig:dipole_col_inviscid}
       \end{figure}

\begin{figure}[!h]
        \centering         
\includegraphics[width=\textwidth]
{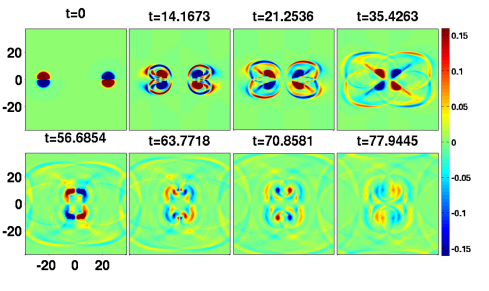}
         \caption{Head to head collision between two equal strength  sub -
luminar dipoles for visco-elastic fluid of ${\Omega_{01}}$=${\Omega_{02}}$=3.5
with the coupling parameters $ {\eta}=5, {\tau_m}=20$.}
               \label{fig:dipole_sub_col_520_zita3p5}
       \end{figure}

\begin{figure}[!h]
        \centering         
\includegraphics[width=\textwidth]
{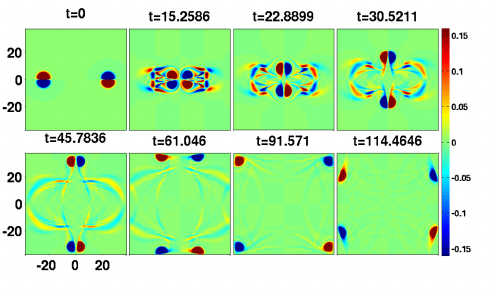}
         \caption{Head to head collision between two equal strength super -
luminar dipoles for visco-elastic fluid of ${\Omega_{01}}$=${\Omega_{02}}$=5
with the coupling parameters $ {\eta}=5, {\tau_m}=20$.}
               \label{fig:dipole_wake_col_520_zita5}
       \end{figure}

\begin{figure}[!h]
        \centering         
\includegraphics[width=\textwidth]
{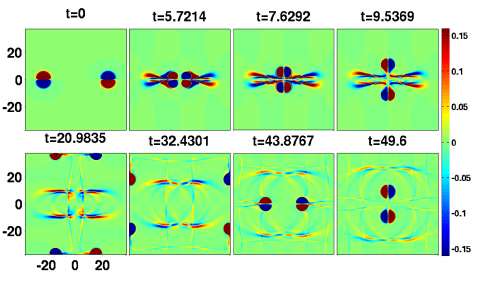}
         \caption{Head to head collision between two equal strength super -
luminar dipoles for visco-elastic fluid each of
${\Omega_{01}}$=${\Omega_{02}}$=10 with the coupling parameters $ {\eta}=5,
{\tau_m}=20$.}
               \label{fig:dipole_sup_col_520_zita10}
       \end{figure}

\paragraph*{}       
%%%%%%%%%%%%%%%%%%%%%%%%%%%%%%%%%%%%%%%%%%%%%%%%%%%%%%%%%%%%%%%%
When  two oppositely propagating dipoles collide with each other, it is well
known in the context of hydrodynamics,  that the their lobes exchange partners
and form new dipolar structure which propagates orthogonal to the initial
propagation direction. This can be observed from the
Fig.~\ref{fig:dipole_col_inviscid}. We consider the two dipoles whose vorticity
profile is given by,
${\xi_{0}(x,y,t_0)}={\xi_{01}(x,y,t_0)}+{\xi_{02}(x,y,t_0)}$. Here, right
side dipolar vorticity is 
${\xi_{01}(x,y,t_0)}={\Omega_{01}}(y-y_{c1})exp\left(-\left({\left(x-x_{c1}
\right)^2+(y-y_{c1})^2}\right)/{a^2_{c1}}\right)$
and left side dipolar vorticity is
${\xi_{02}(x,y,t_0)}={\Omega_{02}}(y-y_{c2})exp\left(-\left({\left(x-x_{c2}
\right)^2+(y-y_{c2})^2}\right)/{a^2_{c2}}\right)$ with parameters
${a_{c1}}$=${a_{c2}}$=2.5, ${x_{c1}}=-24$, ${y_{c1}}$=0, ${x_{c2}}=24$,
${y_{c2}}$=0 and ${\Omega_{01}}$=${\Omega_{02}}$ for equal strength dipoles and
different values have been taken for different cases. In cases of disparate
strength dipoles ${\Omega_{01}} \neq {\Omega_{02}}$.
%%%%%%%%%%%%%%%%%%%%%%%%%%%%%%%%%%%%%%%%%%%%%%%%%%%%%%%%%%%%  
       \FloatBarrier   
%%%%%%%%%%%%%%%%%%%%%%%%%%%%%%%%%%%%%%%%%%%%%%%%%%%%%%%%%%%%  
Similar effect is observed   in the context of collision in i-GHD system. We
again consider the two cases of collision amidst sub and super - luminar pairs
of dipoles in Fig.~\ref{fig:dipole_sub_col_520_zita3p5} and
Figs.~\ref{fig:dipole_wake_col_520_zita5}, ~\ref{fig:dipole_sup_col_520_zita10}
respectively. 

\begin{figure}[!h]
        \centering         
\includegraphics[width=\textwidth]
{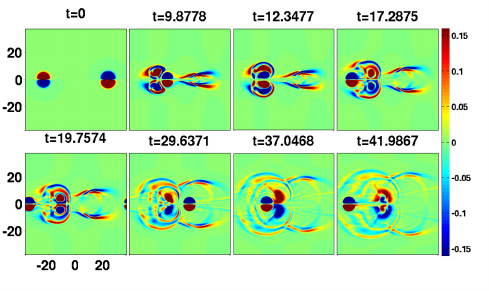}
         \caption{Head to head collision between two disparate strength dipoles,
sub - luminar dipole (left) of ${\Omega_{01}}$= 3.5 and super - luminar dipole
(right) of ${\Omega_{02}}$=10 with the coupling parameters ${\eta}=5,
{\tau_m}=20$.}
               \label{fig:dipole_sub_sup_col_520_zita3p5_10}
       \end{figure}

\begin{figure}[!h]
        \centering         
\includegraphics[width=\textwidth]
{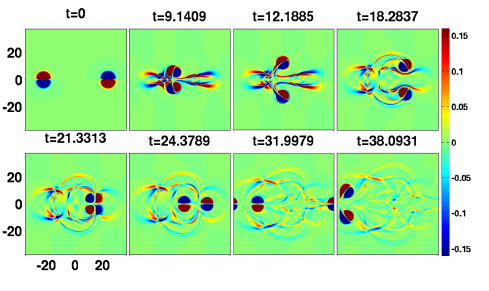}
         \caption{Head to head collision between two disparate
strength super - luminar  dipoles of ${\Omega_{01}}$= 7.5 (left)
and ${\Omega_{02}}$= 10 (right) with the coupling parameters
${\eta}=5, {\tau_m}=20$.}
               \label{fig:dipole_sup_sup_col_520_zita7p5_10}
       \end{figure}

In the former case the radiation engulfs the system. The dipoles slow down
considerably as they move towards each other. This happens due to the preceding
waves from each structure that inhibits their propagation forward. They almost
come to a standstill before they exchange partners and then move in the
orthogonal direction. This can be observed from the
Fig.~\ref{fig:dipole_sub_col_520_zita3p5}. The identity of the dipolar lobes is
ultimately completely lost due to the interaction with the emitted shear waves. 
%%%%%%%%%%%%%%%%%%%%%%%%%%%%%%%%%%%%%%%%%%%%%%%%%%%%%%%%%%%%%  
       \FloatBarrier  
%%%%%%%%%%%%%%%%%%%%%%%%%%%%%%%%%%%%%%%%%%%%%%%%%%%%%%%%%%%%%
In the second case the dipoles exchange partners and move ahead in the
orthogonal direction leaving the radiation behind.
%%%%%%%%%%%%%%%%%%%%%%%%%%%%%%%%%%%%%%%%%%%%%%%%%%%%%%%%%%%%%  
       \FloatBarrier        
%%%%%%%%%%%%%%%%%%%%%%%%%%%%%%%%%%%%%%%%%%%%%%%%%%%%%%%%%%%%%%%%%%%%%%%%%%%%%% 
\paragraph*{}
In Fig.~\ref{fig:dipole_sup_col_520_zita10}, we consider the case of dipoles
approaching toward each other with more strength than  the former cases. The
damage in this case to the lobes is very weak and the dipoles retain their
identity.          
%%%%%%%%%%%%%%%%%%%%%%%%%%%%%%%%%%%%%%%%%%%%%%%%%%%%%%%%%%%%%  
       \FloatBarrier  
%~~~~~~~~~~~~~~~~~~~~~~~~~~~~~~~~~~~~~~~
\paragraph*{}
%~~~~~~~~~~~~~~~~~~~~~~~~~~~~~~~~~~~~~~~
Collisional interactions of disparate strength dipoles have also been studied.
In Fig.~\ref{fig:dipole_sub_sup_col_520_zita3p5_10}, we consider two disparate
strength dipoles. The sub - luminar dipole on left has ${\Omega_{01}}$= 3.5  and
the super - luminar dipole on the right has ${\Omega_{02}}$=10. As opposed to
the normal case where the dipoles of equal strength propagate in the direction
normal to the direction of propagation before collision, for the present case as
evident from Fig.~\ref{fig:dipole_sub_sup_col_520_zita3p5_10}, the the super -
luminar dipole pierces into the lobes of sub - luminar dipole. It can be clearly
seen, after the accomplishment of this crossing process, the lobes of sub -
luminar dipole again come close to each other and start propagating like an
independent dipole. It is interesting to note that there is no exchange of lobes
between dipoles. Both these dipoles (sub and super) propagate in the same
direction as before collision. As the time progresses these dipoles interact
with the wake type structures (left behind by them) and sub - luminar
dipole losses its identity earlier than super - luminar dipole.
%%%%%%%%%%%%%%%%%%%%%%%%%%%%%%%%%%%%%%%%%%%%%%%%%%%%%%%%%%%%%
       \FloatBarrier   
%%%%%%%%%%%%%%%%%%%%%%%%%%%%%%%%%%%%%%%%%%%%%%%%%%%%%%%%%%%%%     
In Fig.~\ref{fig:dipole_sup_sup_col_520_zita7p5_10}, we consider two super -
luminar disparate strength dipoles, of ${\Omega_{01}}$=7.5 (left) and
${\Omega_{02}}$=10 (right). It is observed that, here after exchanging lobes,
these new dipole changes its trajectory and along with the axial motion the
weaker lobe rotates around the stronger lobe. With this rotation, new dipoles of
unequal lobes strength approaches each other and collide again. The exchange of
lobes takes place once again and the newly formed dipoles (with same lobes as
before collision process) starts propagating in the same direction just as
before collision. This collisional process repeats again and again due to
periodic conditions and dipoles also experience the interaction with wake left
behind by them.
%%%%%%%%%%%%%%%%%%%%%%%%%%%%%%%%%%%%%%%%%%%%%%%%%%%%%%%%%%%%%  
       \FloatBarrier  
 %%%%%%%%%%%%%%%%%%%%%%%%%%%%%%%%%%%%%%%%%%%%%%%%%%%%%%%%%%%%%
All these interactions have shown generation of various complicated radiation
and convection patters. We would now take a up a few complex cases and study the
accuracy with which the Poynting like conservation theorem is satisfied.    
%=================================================
\section{Numerical verification of Poynting like Equation for i-GHD}
\label{num_veri_poy_ch2}
\paragraph*{}
We now study the role of different transport processes  in the integral equation
Eq.{~(\ref{eq:integral_equ})} on the evolution of $W$ in the context of monopole
and dipole evolution.
\subsection{For Monopole evolution }
\begin{figure}[!h]
                \centering               
\includegraphics[width=\textwidth]
{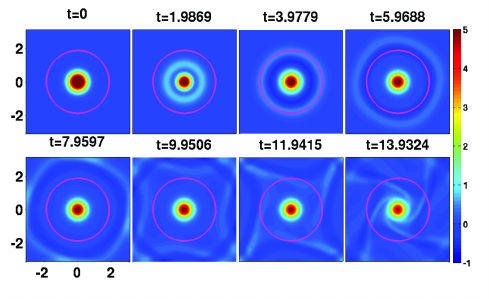}
              \caption{ Evolution of smooth circular vorticity profile in time 
for  visco-elastic fluid with the parameters ${\eta}=2.5, {\tau_m}=20$ and a
circular local volume element (red circumference) over which calculated the
different transport quantities.}
          \label{fig:vort_cnsrv_smooth}
\end{figure}

\begin{figure}[!h]
                \centering               
\includegraphics[width=\textwidth]
{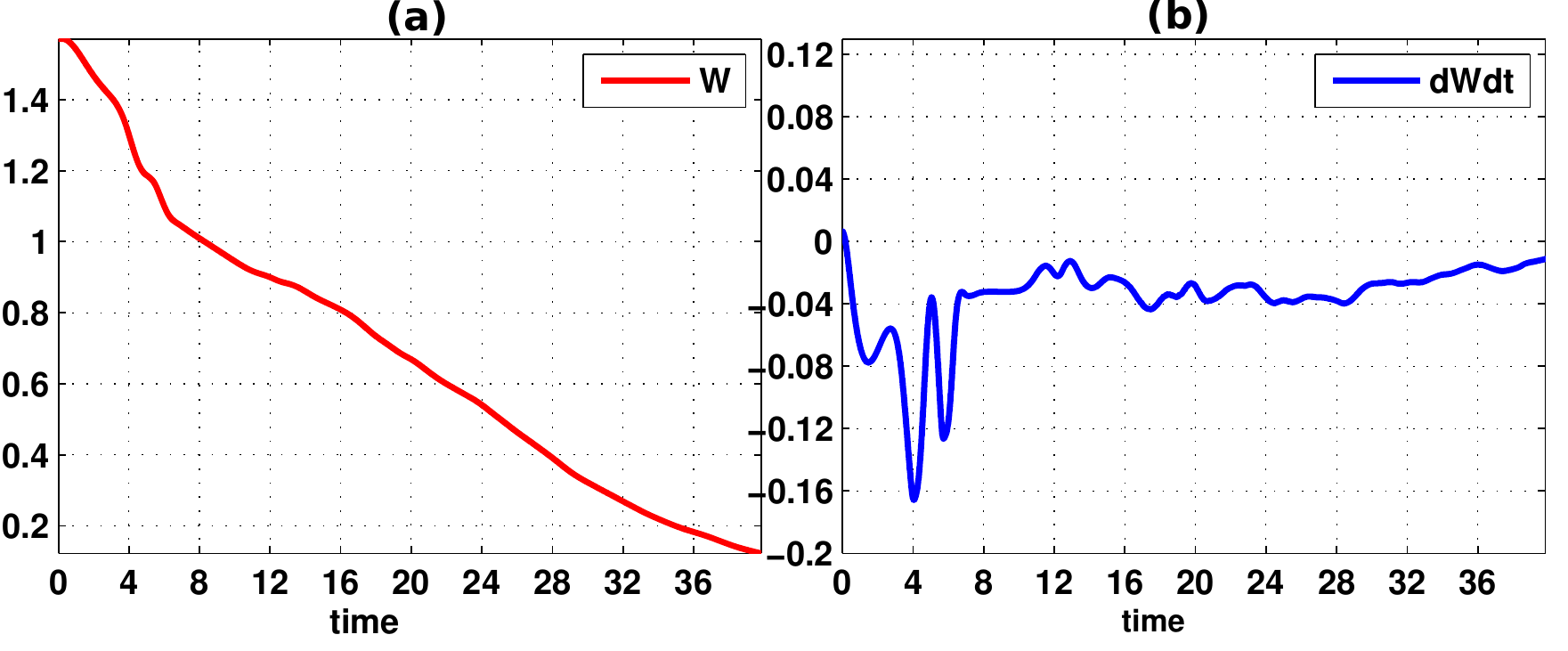}
             \caption{The subplot (a) the conserve quantity $W$ within our
considered regime with time and the subplot (b) is the  slope of  $W$}
          \label{fig:conservation_w_smooth}
\end{figure}

\begin{figure}[!h]
                \centering               
\includegraphics[width=\textwidth]
{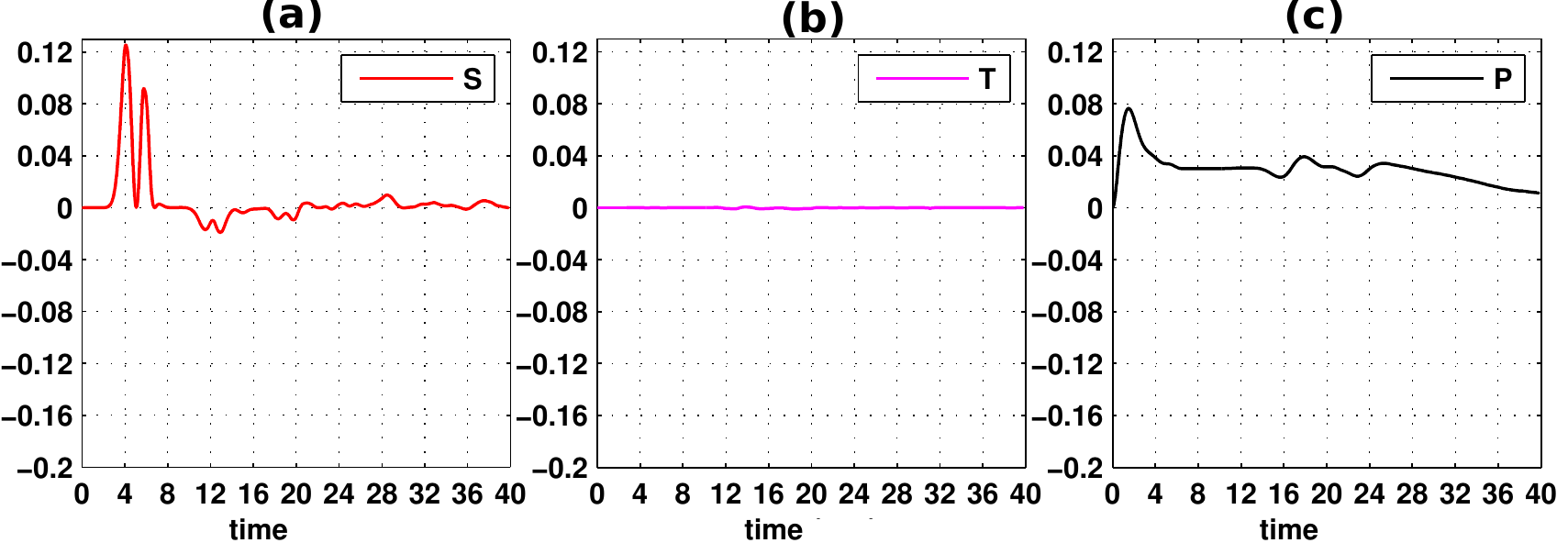}
               \caption{The contribution of each quantity i.e. (a) S, (b) T and
(c) $P_d$ }
          \label{fig:conervation_all_smooth}
\end{figure}

\begin{figure}[!h]
                \centering               
\includegraphics
[height=10.0cm,width=\textwidth]
{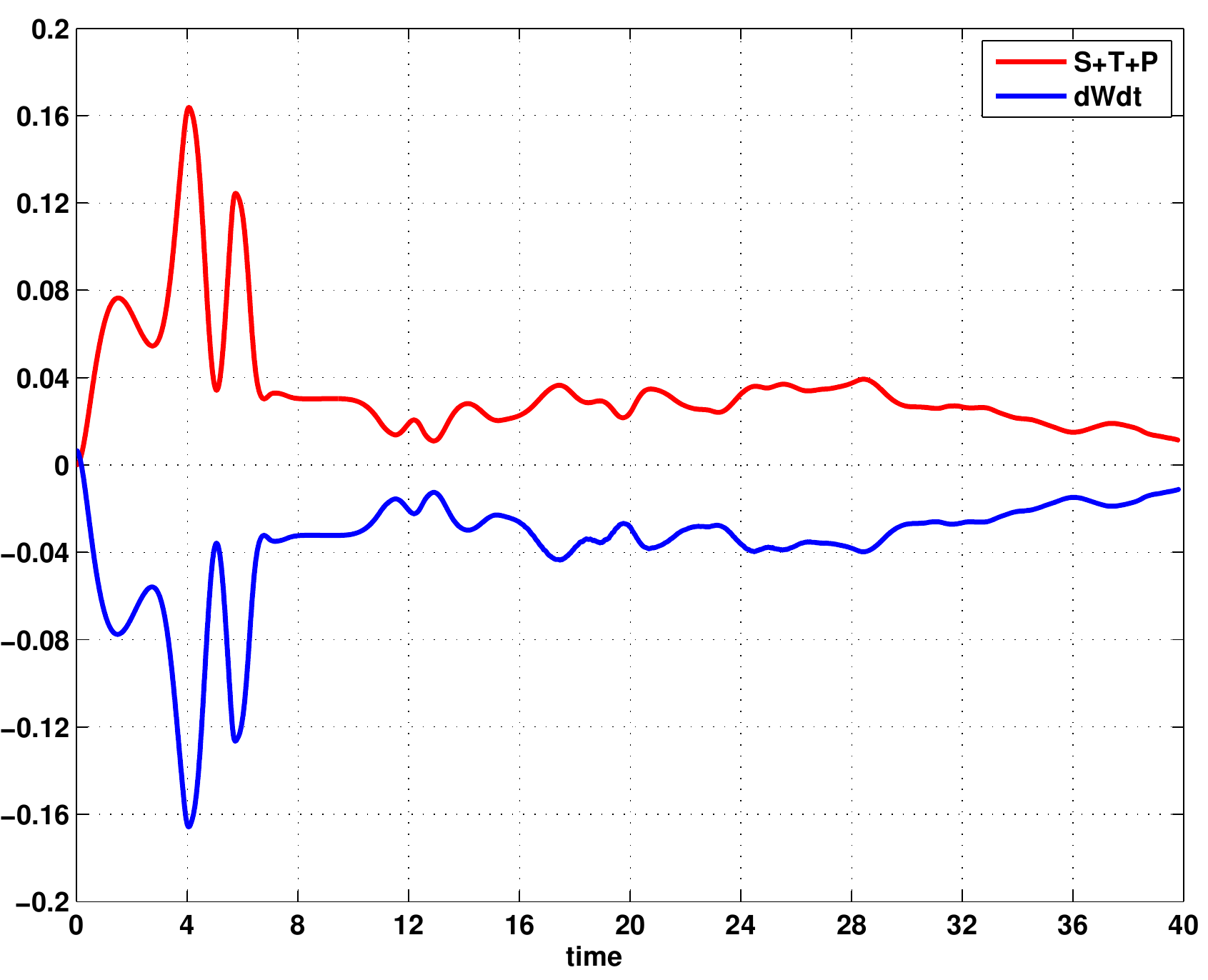}
               \caption{ Time derivative of conserved quantity $W$ ({\bf
$\color{blue}{ -} $}) is the mirror image to the total sum (S+T+P) of all
remaining quantities ({\bf $\color{red}{-} $}) during the run time}
          \label{fig:conservation_ref_smooth}
\end{figure}

\paragraph*{}
 We investigated the emission of transverse shear waves from the rotating smooth
vorticity profile in strongly coupled dusty plasma medium \cite{Dharodi_2014}.
In this case the vorticity smooth profile is given by,
${\xi_{0}(x,y,t_0)}={\Omega_0}exp\left(-\left({\left(x-x_c
\right)^2+(y-y_c)^2}\right)/{a^2_c}\right)$. The numerical simulation has been
carried out for ${a_c}$=0.5, ${\Omega_0}=8$ and ${x_c}={y_c}$=0. We found that
phase velocity $v_p$ of such waves is proportional to the coupling strength of
the medium. In Fig.~\ref{fig:vort_cnsrv_smooth} the evolution of a circular
vorticity patch in the strong coupling limit with parameters (${\eta}=2.5;
{\tau_m}=20$) for GHD system has been shown. A red circle with a radius of
0.6$\pi$ has been drawn in the plots. Initially all the action is within this
circular boundary. However, as time progresses the waves are emitted which cross
this boundary. We investigate the validity of the integral
Eq.{~(\ref{eq:integral_equ})} within this boundary. Our simulation region is a
square box of length $2 \pi$ units with periodic conditions. The periodic
condition ensures that the waves would not only propagate out of the circular
demarcated region but would also enter it subsequently from the other side due
to the periodicity of the square box. In fact the evidence is clear from the
subplots in the second row of the Fig.~\ref{fig:vort_cnsrv_smooth}.
 %%%%%%%%%%%%%%%%%
\FloatBarrier
% %%%%%%%%%%
 The change in the magnitude of  $W$ within the circular region regime with time
is shown in subplot (a) of Fig.~\ref{fig:conservation_w_smooth}. We observe a
steady decay in magnitude of $W$. It is clear from the plot that the rate of
decay of $W$ is not a constant. Thus, the sum of  contribution of various terms
in Eq.{~(\ref{eq:integral_equ})}  which defines the evolution  of $W$ changes
with time. The subplot Fig.~\ref{fig:conservation_w_smooth} (b) is $dW/dt$  the
slope of $W$ shown in  Fig.~\ref{fig:conservation_w_smooth} (a). The evolution
of various terms has been shown in the subplots of 
Fig.~\ref{fig:conervation_all_smooth}.
%%%%%%%%%%%%%%%%%%%%%%%%%%%%%%%%%%%%%%%%%%%%%%%%%%%%
\FloatBarrier
%%%%%%%%%%%%%%%%%%
 The subplot (Fig.~\ref{fig:conervation_all_smooth} (a)) represents the change
in $W$ by wave emission. It is positive when the waves leave the region and
negative when they enter the region.  The comparison of subplot
(Fig.~\ref{fig:conervation_all_smooth} (a)) with
Fig.~\ref{fig:vort_cnsrv_smooth}, one can clearly say that the positive peak 
in this subplot corresponds to the time when the transverse shear waves pulse 
leaves the circular patch. Similarly,  the negative peak here denotes the 
time when the waves enter the region after re-entering the simulation box 
from the other end due to the periodic boundary condition. Since, the  monopolar
vortex remains stationary and merely rotates about its axis so there is no 
convection of the fluid across the region. Thus,  the subplot
(Fig.~\ref{fig:conervation_all_smooth} (b)) shows no contribution. The role of
dissipating term is shown in subplot of Fig.~\ref{fig:conervation_all_smooth}
(c), which is also observed to be finite.  
%~~~~~~~~~~~~~~~~~~~~~~~~~~~~~~~~~~~~~~~
\paragraph*{}
%~~~~~~~~~~~~~~~~~~~~~~~~~~~~~~~~~~~~~~~
It should be noted that while the contribution from the Poynting flux of 
wave and convective term can either decrease or increase $W$, the last
dissipative terms is always positive and would only cause $W$ to decay. 
%~~~~~~~~~~~~~~~~~~~~~~~~~~~~~~~~~~~~~~~~~~~~~~~~~
\FloatBarrier
%~~~~~~~~~~~~~~~~~~~~~~~~~~~~~~~~~~~~~~~~~~~~`
In Figure~\ref{fig:conservation_ref_smooth} we plot $dW/dt $({\bf 
$\color{blue}{ -} $}) and the sum of all the three terms (S+T+P)  ({\bf $\color{red}{ -} $})
 separately. It can be seen that the two plots are the 
an accurate mirror image of each other proving that their sum is exactly zero 
as expected from Eq.{~(\ref{eq:integral_equ})}.
%%%%%%%%%%%%%%%%%%
\FloatBarrier
\subsection{For Dipole evolution}
\begin{figure}[!h]
                \centering               
\includegraphics[width=\textwidth]
{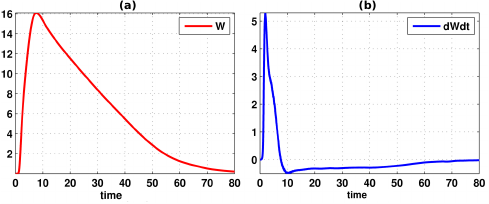}
  \caption{The subplot (a) the conserve quantity $W$ within our considered
regime with time and the subplot (b) is the  slope of  $W$.}
          \label{fig:dipole_w_520_zita3p5}
\end{figure}

\begin{figure}[!h]
                \centering               
\includegraphics[width=\textwidth]
{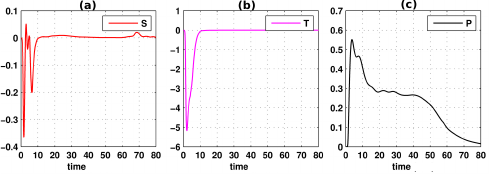}
               \caption { The subplot (a) represents the change in $W$ by wave
emission. In case of dipole the major transport process is due to the convection
phenomena can be seen in subplot (b). The role of dissipating term is shown in
subplot (c), which is observed to be finite.}
          \label{fig:dipole_all_520_zita3p5}
\end{figure}

\begin{figure}[!h]
                \centering               
\includegraphics
[height=10.0cm,width=\textwidth]
{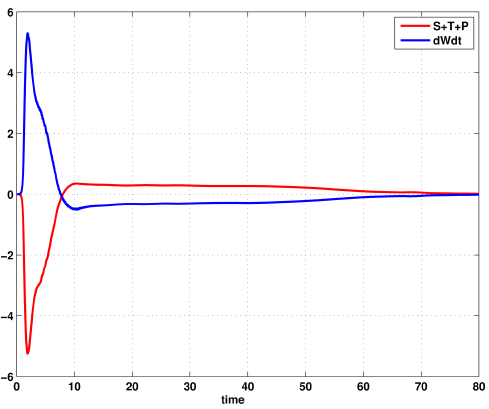}
               \caption{ Time derivative of conserved quantity $W$ ({\bf
$\color{blue}{ -} $}) is the mirror image to the total sum (S+T+P) of all
remaining quantities ({\bf $\color{red}{ -} $}) during
the run time.}
          \label{fig:dipole_ref_520_zita3p5}
\end{figure}

\begin{figure}[!h]
                \centering               
\includegraphics[width=\textwidth]
{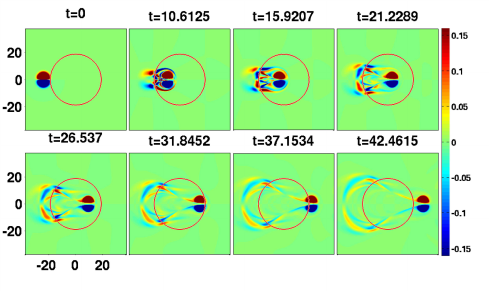}
              \caption{ Evolution of dipole with time for
visco-elastic fluid of ${\Omega_0}$=5.0 with the coupling parameters ${\eta}=5,
{\tau_m}=20$ and a circular local volume element (red circumference) over which
calculated the different transport quantities.}
          \label{fig:dipole_520_vort_zita5_con}
\end{figure}

\begin{figure}[!h]
                \centering               
\includegraphics[width=\textwidth]
{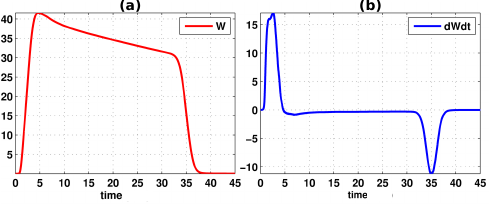}
 \caption{The subplot (a) the conserve quantity $W$ within our considered
regime with time and the subplot (b) is the  slope of  $W$}
          \label{fig:dipole_w_520_zita5}
\end{figure}

\begin{figure}[!h]
                \centering               
\includegraphics[width=\textwidth]
{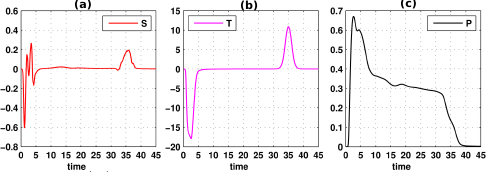}
             \caption { The subplot (a) represents the change in $W$ by wave
emission. In this case also the major transport process is due to the convection
phenomena can be seen in subplot (b). The role of dissipating term is shown in
subplot (c), which is observed to be finite.}
          \label{fig:dipole_all_520_zita5}
\end{figure}

\begin{figure}[!h]
                \centering               
\includegraphics
[height=10.0cm,width=\textwidth]
{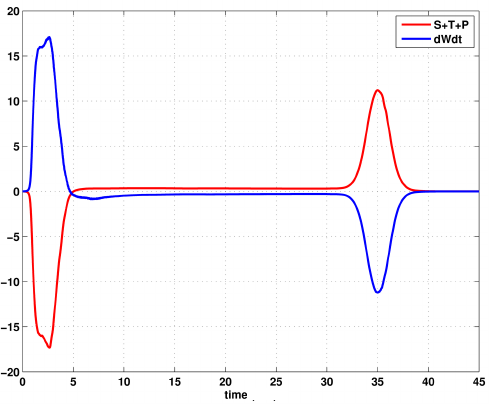}
               \caption{ Time derivative of conserved quantity $W$ ({\bf
$\color{blue}{ -} $}) is the mirror image to the total sum (S+T+P) of all
remaining quantities ({\bf $\color{red}{ -} $}) during
the run time.}
          \label{fig:dipole_ref_520_zita5}
\end{figure}

\begin{figure}[!h]
                \centering               
\includegraphics[width=\textwidth]
{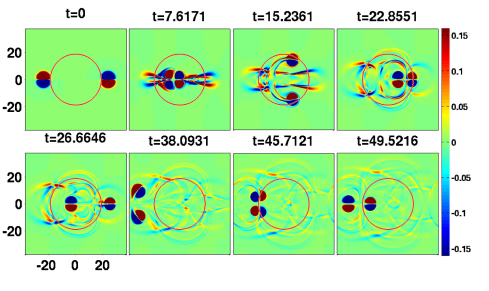}
            \caption{ Head to head collision between two disparate
strength super - luminar  dipoles of ${\Omega_{01}}$= 7.5 (left)
and ${\Omega_{02}}$= 10 (right) with the coupling parameters
${\eta}=5, {\tau_m}=20$ and a circular local volume element (red circumference)
over which calculated the different transport quantities.}
          \label{fig:dipole_vort_sup_sup_col_520_zita7p5_10_con}
\end{figure}

\begin{figure}[!h]
                \centering               
\includegraphics[width=\textwidth]
{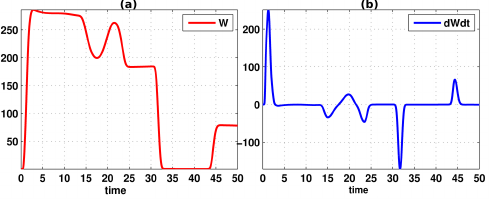}
 \caption{The subplot (a) the conserve quantity $W$ within our considered
regime with time and the subplot (b) is the  slope of  $W$}
          \label{fig:dipole_col_w_520_zita7p5_10}
\end{figure}

\begin{figure}[!h]
                \centering               
\includegraphics[width=\textwidth]
{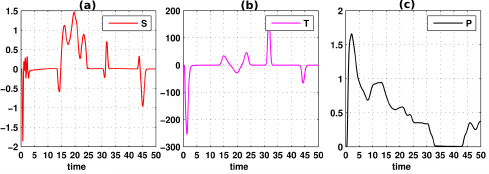}
             \caption { The subplot (a) represents the change
in $W$ by wave emission. In this case also the major transport process
is due to the convection phenomena can be seen in subplot (b). The role of
dissipating term is shown in subplot (c), which is observed to be finite.}
          \label{fig:dipole_col_all_520_zita7p5_10}
\end{figure}

\begin{figure}[!h]
                \centering               
\includegraphics
[height=10.0cm,width=\textwidth]
{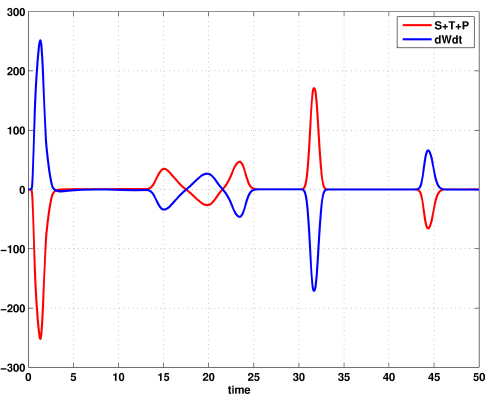}
               \caption{ Time derivative of conserved quantity $W$ ({\bf
$\color{blue}{ -} $}) is the mirror image to the total sum (S+T+P) of all
remaining quantities ({\bf $\color{red}{ -} $}) during
the run time.}
          \label{fig:dipole_col_ref_520_zita7p5_10}
\end{figure}

\paragraph*{}
Monopoles being static structures, the contribution due to the convective terms
in the equation was negligible as we saw in the previous subsection. We now 
choose some specific cases of dipoles evolution from earlier
section{~(\ref{dipole_evo_ch2})} and study the evolution of the various
terms in the Eq.{~(\ref{eq:integral_equ})} in a circular region of radius $6\pi$. 
 Again, simulation region is a square box of length $ 24\pi$ units with
periodic conditions. The periodic condition ensures that the dipole as well as
the emitted waves can enter and leave the region multiple times. The system
parameters (system length and circular local volume element) and coupling
parameters ($\eta=5$ and $\tau_m=20$) are same for all the  cases mentioned
below. We now show the validity of Eq.{~(\ref{eq:integral_equ})} for different
dipoles with varying strength, $\Omega_0$, in the subsequent discussion .
% We are going to present that the  Eq.{~(\ref{eq:integral_equ})}
% holds good for the dipole evolution and their collision with different strength
% parameter $\Omega_0$. 
 %%%%%%%%%%%%%%%%%%
        \begin{figure}[!h]
                \centering               
\includegraphics[width=\textwidth]
{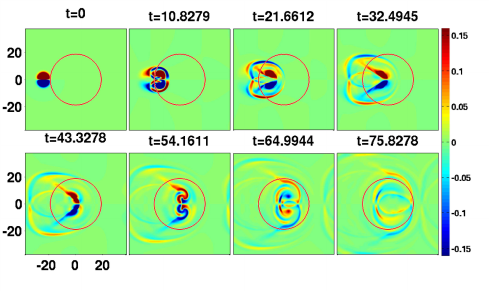}
             \caption{ Evolution of dipole with time for
visco-elastic fluid of ${\Omega_0}$=3.5 with the coupling parameters ${\eta}=5,
{\tau_m}=20$ and a circular local volume element (red circumference) over which
calculated the different transport quantities.}
          \label{fig:dipole_520_vort_zita3p5_con}
\end{figure}
\FloatBarrier
% %%%%%%%%%%
Fig.~\ref{fig:dipole_520_vort_zita3p5_con} shows the propagation of the dipolar
structures along with the emitted waves. The region inside the red circle is
considered for studying the evolution of $W$. 
 %%%%%%%%%%%%%%%%%%%%%%%%%%%%%%%%%%%%%%%%%%%%%%%%%%%%
 The total change in magnitude of conserved quantity $W$ within this region with
time is shown in Fig.~\ref{fig:dipole_w_520_zita3p5} (a). Since the dipole was
placed outside this region initially $W$ was zero to begin with. When the dipole
enters this boundary at around time 1.0 $W$ shows a sharp increase. As time
progresses the value of $W$ steadily falls owing to the dissipative term
(cf. Fig.~\ref{fig:dipole_all_520_zita3p5}(c)).
 %%%%%%%%%%%%%%%%%%%%%%%%%%%%%%%%%%%%%%%%%%%%%%%%%%%%
\FloatBarrier
%~~~~~~~~~~~~~~~~~~~~~~~~~~~~~~~~~~~~~~~~~~~~~~~~~~~~~~~~~~~~~
When dipole enters this region at around time 1.0, the contribution of
transverse and the convection terms can be seen clearly in
subplots ~\ref{fig:dipole_all_520_zita3p5}(a),
~\ref{fig:dipole_all_520_zita3p5}(b)
respectively.
%%%%%%%%%%%%%%%%%%
\FloatBarrier
%~~~~~~~~~~~~~~~~~~~~~~~~~~~~~~~~~~~~~~~~~~~~
The conservation equation is pretty accurately satisfied as can be seen from
Figure~\ref{fig:dipole_ref_520_zita3p5} where $dW/dt$ (blue line) and  the sum
of the three terms has been plotted by red line. They are identical mirror image
curves illustrating that the conservation equation $dW/dt +S+P+T = 0$ is
satisfied with very good precision.
%%%%%%%%%%%%%%%%%%
\FloatBarrier
%=====================================================================
In earlier case (Fig.~\ref{fig:dipole_520_vort_zita3p5_con}), the dipole was of
lesser strength and hence it dissipated inside the  circular region considered by us.
 In Fig.~\ref{fig:dipole_520_vort_zita5_con}, the strength of the dipole is
chosen to be sufficiently high so that it can cross the region marked by the red
circle over which we are calculating different transport quantities.
%========================================================================
\FloatBarrier
%=====================================================================
From  Fig.~\ref{fig:dipole_520_vort_zita5_con}, it is clear that as the
dipole enters and leaves the considered circular region at around time  $1$ and
$35$ respectively, there is sharp rise and fall in $W$ is also observed as
shown in subplot~\ref{fig:dipole_w_520_zita5} (a). In this time duration the
convection term contribute significantly as seen in subplots
~\ref{fig:dipole_all_520_zita5}(b) and the contribution of radiation  term is
smaller as compared to the convection term
(cf. subplot~\ref{fig:dipole_all_520_zita5}(a)). However, in the intervening
time a steady decrease in $W$ occurs mainly because of the dissipative term 
shown in subplot~\ref{fig:dipole_all_520_zita5}(c).
 %%%%%%%%%%%%%%%%%%%%%%%%%%%%%%%%%%%%%%%%%%%%%%%%%%%%
\FloatBarrier
 %%%%%%%%%%%%%%%%%%
\FloatBarrier
From Fig.~\ref{fig:dipole_ref_520_zita5}, one observes that $dW/dt$ is the
mirror image to the total sum (S+T+P) of all remaining quantities during the run
time as observed for earlier cases.
%%%%%%%%%%%%%%%%%%
\FloatBarrier
%%%%%%%%%%%%%%%%%%%%%%%%%%%%%%%%%%%%%%%%%%%%%%%%%%%%
{\bf{Collision}}. 
\paragraph*{}
In order to confirm the validity of the conservation relation for more complex
scenario, we consider the case of two colliding super - luminar
disparate strength dipoles of ${\Omega_{01}}$=7.5 (left) and ${\Omega_{02}}$=10
(right) shown in Fig.~\ref{fig:dipole_vort_sup_sup_col_520_zita7p5_10_con}.
 %%%%%%%%%%%%%%%%%%%%%%%%%%%%%%%%%%%%%%%%%%%%%%%%%%%%
\FloatBarrier
%~~~~~~~~~~~~~~~~~~~~~~~~~~~~~~~~~~~~~~~~~~~~~~~~~~~~~~~~~~~~~~~~~~~~~~~
The complexity of the motion of dipole is evident in 
Fig.~\ref{fig:dipole_vort_sup_sup_col_520_zita7p5_10_con}. Here the
dipoles exhibits linear/circular motion and collides multiple times inside,
outside and along the circumference. This complex evolution of dipoles can be
intimately related to the evolution of various terms in the conservation
relation. In time duration $5$ to $10$, the dipoles collides axially inside the
considered region and then move along the circumference at around time $14$.
During this circular motion dipoles share inside and outside vary and finally
collide orthogonally to the first collision. This is well reflected in the
fluctuation in plot of $W$ shown in
Fig.~\ref{fig:dipole_col_w_520_zita7p5_10}(a). Further, in time range ($33$ to
$43$) dipoles leave completely this region so $W$ almost becomes zero. Again we
observe a sharp increase in $W$ at time around $44$ because of collision between
dipoles at the circumference.
 %%%%%%%%%%%%%%%%%%%%%%%%%%%%%%%%%%%%%%%%%%%%%%%%%%%%
\FloatBarrier
%~~~~~~~~~~~~~~~~~~~~~~~~~~~~~~~~~~~~~~~~~~~
 These  events can be corroborated well by observing the contour plot of
 Fig.~\ref{fig:dipole_vort_sup_sup_col_520_zita7p5_10_con} and the  evolution of
the various  terms namely, the convective, Poynting and dissipative terms shown in
Fig.~\ref{fig:dipole_col_all_520_zita7p5_10}.
% The contribution of each individual term which occurs due to the emission of
% waves,convection and dissipation term shown in subplots of
% Fig.~\ref{fig:dipole_col_all_520_zita7p5_10} (a), (b) and (c) respectively.
 %%%%%%%%%%%%%%%%%%
\FloatBarrier
Here too the integral condition of Eq.{~(\ref{eq:integral_equ})}
is satisfied identically at all times it can be seen
from Fig.~\ref{fig:dipole_col_ref_520_zita7p5_10}
%%%%%%%%%%%%%%%%%%
\FloatBarrier
%%%%%%%%%%%%%%%%%%%%%%%%%%%%%%%%%%%%%%%
% The study of monopole and dipolar structures made thus far reveals the fact
% that the conservation relation can be formulated for systems governed by the
% GHD equation and that complex evolution of the system can be understood in
% terms of the 
% Thus, from all above cases (monopole and dipolar structures) one can find out
% the contribution of each individual term which occurs due to the transverse
% waves, convection and dissipation in system.
%~~~~~~~~~~~~~~~~~~~~~~~~~~~~~~~~~~~~~~~~~~~
\section{Summary and conclusion}
\label{discussionch2}
\paragraph*{}
The evolution of smooth vorticity profile for a strongly coupled medium by the
visco-elastic GHD description have been studied. In contrast to the Newtonian
fluid the GHD visco-elastic medium, we have shown numerically the emission of
transverse shear waves traveling  with phase velocity
${\sqrt{{\eta{}}/{\tau_m}}}$ as expected analytically from GHD model. The
propagating dipole structures have been studied extensively in both  sub/super
luminar limits (i.e. when the propagation speed of the dipole is 
slower/faster than the  TSW phase velocity). In the sub - luminar case the
dipole remains confined inside the radiation that it emits. Thus the radiation
continuously backreacts on the structure which looses its identity soon. On the
other hand the radiation emitted by the super - luminar dipole remains confined
in a conical wake region of the structure and is unable to distort it. The
dipole thus continues to maintain its identity for a much longer duration.
Detailed collision studies amidst these structures where also carried out.

A Poynting like conservation theorem was also constructed. The theorem was shown
to be satisfied for the GHD set of equations pretty accurately for many complex 
evolution cases.
% ---------------------------------------------------------
% \newpage
 \bibliographystyle{unsrt}
 \bibliography{vikram_conserve}
%~~~~~~~~~~~~~~~~~~~~~~~~~~~~~~~~~~~~~~~~~~~~~~~~~~~~~~~~~~
\end{document}